# Strain-induced effects in the electronic and spin properties of a monolayer of ferromagnetic GdAg$_2$

A. Correa,[1,2,3] B. Xu,[4,5,6] M. J. Verstraete[4,5] and L.Vitali[1,3,7]

[1] Departamento de física de materiales, Universidad del País Vasco, 20018 San Sebastian (Spain)

[2] Donostia International Physics Center, 20018 San Sebastian (Spain)

3 Centro de Fisica de Materiales y Material Physics Center, 20018 San Sebastian (Spain)

[4] CESAM and Département de Physique, Université de Liège, B5a Allée du 6 Août 19, B-4000 Sart Tilman, Belgium

[5] European Theoretical Spectroscopy Facility (http://www.etsf.eu)

[6] Department of Physics and Institute for Nanoscience and Engineering, University of Arkansas, Fayetteville, Arkansas 72701, USA

[7] Ikerbasque Foundation for Science, 48013 Bilbao (Spain)

**We report on the structural, electronic and magnetic properties of a monolayer of GdAg$_2$, forming a moiré pattern on Ag(111). Combining scanning tunneling microscopy and *ab-initio* spin-polarized calculations, we show that the electronic band structure can be shifted linearly via thermal dependent strain of the intra-layer atomic distance in a range between 1-7%, leading to lateral hetero-structuring. Furthermore, the coupling of the incommensurable GdAg$_2$ alloy layer to the Ag(111) substrate leads to spatially varying atomic relaxation causing subsurface layer buckling, texturing of the electronic and spin properties, and inhomogeneity of the magnetic anisotropy energy across the layer. These results provide perspectives for a control of electronic properties and magnetic ordering in atomically-thin layers.**

Geometrical structure, lattice periodicity and atomic arrangement are subtly intertwined with the electronic properties of materials. Sub-angstrom changes in the atomic distance are sufficient to modify the physical and chemical properties, such as the band-structure, carrier mobility and the chemical reactivity[1-15]. The deposition of two-dimensional layered crystals on mechanically stretchable or bendable substrates can produce one-dimensional strained structures with intriguing properties[1,2]. Similarly, biaxial-strained two-dimensional layers have been obtained exploiting the interface energy between lattice mismatched epitaxial layers[3-9,15]. The efficiency of these strategies has generated considerable progress in tailoring the electronic and optical properties[3,8-12,15]. Few sparse works have demonstrated instead the role of surface-strain in bulk or thick layers of ferromagnetic materials [4-7,12-14]. On the other hand, in mesoscopic systems formed by few-layers thick ferromagnetic materials, structure relaxation processes lead to interesting effects in the electronic and magnetic properties [15]. This suggests strain as a good method to manipulate the electronic properties, and consequently the magnetic order, also of thin layers of strong magnetic materials. Nonetheless, a general picture on the influence of atomistic structure on the local magnetic order is still missing.

Here, we will show that the local atomic arrangement critically affects the electronic properties of two-dimensional layer and predict their effect on the magnetic properties of the system. Surface strain inherent in epitaxially-grown two-dimensional layers is sufficient to texture the electronic and the magnetic properties of a ferromagnetic structure stable up to 85K [16]. Specifically, we will characterize a monolayer structure based on rare-earth surface stoichiometric alloy $GdAg_2$ forming a weakly interacting moiré superstructure on Ag(111). This monolayer is particularly suitable to investigate the relation between the structural, electronic and magnetic properties. Indeed, in rare-earth based materials, the exchange interaction process and, consequently, the magnetic order are critically sensible to structural variations and orbital hybridizations [5-7, 12-14] which affects the delicate interplay between the 5d, 6s nearly-free conduction and the highly-localized 4f electrons.

Here, we will demonstrate that in-plane lattice strain conveys to the formation of hetero-structures with variable alloy unit-cell (in the range of 5.15-5.6Å) and distinct density of states. Moreover, the incommensurate alloy layer induces an out-of-plane atomic buckling of the supporting substrate with the periodicity of the moiré structure. This prompts a position dependent coupling interaction to the Ag(111) surface, steers a modulation in the density of states and leads to a local weakening of the ferromagnetic order resulting in a spin-texture across the layer. The direct comparison of scanning tunneling microscopy (STM) and spectroscopy (STS), with state-of-the-art spin-polarized numerical simulations based on density functional theory (DFT), facilitate the understanding of the role of strain, structural relaxation and coupling interaction on the physical properties of the system.

A monolayer of $GdAg_2$ forms an incommensurate structure on Ag(111), similar to the previously reported $GdAu_2$ on the Au(111) surface,[16,17] (see Electronic Supplementary Information ESI1). This results in topographic STM images (Figure 1a) as a moiré superstructure, whose apparent minima and maxima reflect the variation of the geometrical registry of the atoms of the alloy layer with respect to the Ag(111) surface. Three of the canonical geometrical configurations, namely *hcp*, *fcc* and *top*, through which the atoms of the continuous alloy layer must pass periodically are shown in Figure 1b. Considering each of these stacking configurations separately, we provide a picture of the structural and electronic properties across the alloy layer and at its interface with Ag(111).

Using first-principles calculations[18-19] (details in Electronic Supplementary Information ESI2), we show that the formation of the superstructure induces different relaxations of the supporting substrate according to the stacking registry of the alloy layer. The Gd atoms are found 2.872 Å (2.870 Å) above the substrate in the fcc (hcp) stacking configuration, while for the top configuration the preferred position is slightly lower, i.e. 2.794 Å. Correspondingly, the Ag atoms of the substrate below the Gd atoms are pushed into the bulk of the crystal (Figure 1b). The modeling of these configurations allows us to assign the top stacking to the valleys of the topographic image and the fcc and hcp to the hills (i.e., to the "dark" and "bright" sites), respectively. The clearly distinct relaxation of the system in top stacking has a critical role in the electronic properties and in the local spin order of the alloy, as will be shown here.

A closer analysis of Figure 1a shows the presence of two neighboring moiré superstructures differing in their relative orientation and periodicity where the upper-right one has a large moiré lattice constant (hereafter, LMLC) and the other a smaller one (SMLC). The local density of states, measured using standard lock-in techniques on the hills of the two super-structures, is shown as red and blue lines in Figure 1c. The corresponding dI/dV spectra achieved on the valley positions and space-resolved energy maps are shown in Electronic Supplementary Information ESI3-ESI5. The local spectroscopic data shown in Figure 1 demonstrate clearly that the two superstructures differ both in their occupied as well as the empty states. A characteristic feature at 400meV, visible as a shoulder of a higher energy peak occurring at about 630meV, characterizes the LMLC pattern. This peak (hereafter, labeled "X") shifts to higher energy (710 meV) as the lattice constant of the moiré pattern is reduced. An unequivocal localization of these states in either of the two moiré areas can be evinced by the conductance maps (Figure 1 c and d), corroborating the formation of lateral electronic hetero-structures. Further differences between the two superstructures can be observed at lower-energy, where an additional small peak can be observed at 80meV and at higher- energy (see Electronic Supplementary Information ESI3-ESI4). Also the occupied density of state differs in the two superstructures. Their reduced intensity and many shoulders impede, however, a clear quantification of their energy positions. The overall emerging picture is that a reorganization of the electronic properties takes place as a function of the superstructure lattice size. Notably, the surface state of Ag(111) cannot be further observed through the moiré pattern. This suggests that a modification of the electronic properties of the noble metal surface has occurred despite the formation of moiré patterns is indicative of a weak coupling to the supporting substrate. The spectra shown in Figure 1b are representative of a series of dI/dV measurements achieved on different moiré superstructures. A careful data analysis of these shows that the energy position of peak X varies as a function of the lattice constant of the superstructure according to the statistical distribution shown in Figure 2. The observed energy values cluster around two mean values of the superstructure periodicity, 32 and 34Å. The close relation between the electronic structure and lattice constant is further corroborated by a fit of the data (blue line in Figure 2), which suggests a linear dependence. This is confirmed by first principle calculations, as it will be shown in the following.

To understand the observed correlation of the lattice constant of the superstructure and the electronic properties, we have applied to the system the coincidence model proposed by Hermann to describe moiré patterns[21]. This model envisions the moiré super-structure as a coincidence network formed by the atoms of the overlayer that periodically match those of the substrate. Outcomes of the model are the size of the unit cell of the alloy and its angle of rotation α with respect to the Ag(111) high symmetry directions (see Electronic Supplementary Information ESI6 for further details). The structural parameters of two moiré patterns among all predictable for this system, (table 1) allow a direct correspondence to the experimentally observed high-resolution images on the two moiré superstructures (see Electronic Supplementary Information ESI3). Experimental results and modeling consistently show that the unit cell size of the alloy monolayer and its orientation differ in the two moiré patterns seen in in Figure 1. The two periodicities of 32 and 34 Å respectively, correspond to alloy unit cells differing in size by a relative strain of 1%, while the angle α is almost constant. The superstructure with larger periodicity, i.e. with the larger coincidence distance, is formed by the alloy layer with the smaller lattice constant, in good agreement with the experimental observations[21-23] and intuitive expectations.

Guided by this understanding of the structural origin of the two moiré patterns, we simulated their electronic properties with density functional theory (see Electronic Supplementary Information ESI2). The band structure of the GdAg$_2$ alloy is first calculated with the supporting layer Ag(111) and then compared with the free standing case. In this way, we separate the effect of strain on the alloy from the interaction with the substrate, achieve a base line for the band character, estimate the importance of surface interactions, and of the magnetic anisotropy energy. The electronic structure of the GdAg$_2$/Ag(111) was calculated using a commensurate √3x√3 supercell of the noble metal (111) surface in the plane in fcc stacking, which is the configuration with lowest-energy (Figure 3a). The hcp and fcc stackings (Electronic Supplementary Information ESI2-ESI6-ESI7 share most properties, but differ from the top case in the band energy positions and dispersion.

The free-standing $GdAg_2$ alloy layer, calculated using the same parameters as in the supported case, is shown in Figure 3b. Upon the removal of the supporting substrate few bands shift towards lower energies considerably more than the other bands of the system (C and C´). This is an artefact of the free-standing approximation, which highlights their coupling potential with the Ag(111) surface. The effective coupling potential acting on the surface-supported alloy monolayer is expected to be intermediate between the free standing and the theoretical commensurate case shown in Figure 3a and b, respectively.

The free standing alloy is then progressively strained to survey the impact on the band structure. Relaxed and strained structures are compared in Figure 3b and c. The color code of the lines highlights in both cases the correspondence of the two bands around the Γ point, namely X and D, whose contribution dominates the dI/dV spectra. These have a mixed Gd-d, Ag p and Ag d character and play a critical role in the magnetic character of the system[16]. Outcome of these calculations is the relative position of the D and X bands as a function of strain, while their absolute energy position is shifted towards negative energies by the approximation used as discussed above. By increasing the size of the alloy unit cell, i.e., decreasing the periodicity of the moiré supercell, the empty states band X shifts towards higher energies (Figure 3c) explaining quite straightforwardly the shift of peak X shown in Figure 1 and 2. A similar effect was observed also in graphene layers and in $MoS_2$ where a linearly increasing gap of about 100meV per strain percentage was reported[24-25]. Energy shifts are also predicted for the occupied states (red line). The clearly diminishing intensity at low energy and the appearance of a peak at 80meV with the increased lattice size confirm the predicted trend of an upward shift of the D band. Bearing in mind the shortcomings of the used approximation and the onset of Ag(111) bulk state[27], not visible in the present calculations, the agreement between experimental and theoretical results can explain quite straightforwardly the main feature of the density of states as a function of lattice structure.

Remarkably, on Au(111) the $GdAu_2$ alloy do not form moiré hetero-structures. This can be explained considering the calculated surface binding energies using $E_{surf} = E_{hetero} - E_{ML} - E_{slab}$, where $E_{hetero}$, $E_{ML}$, and $E_{slab}$ are the total energies of the hetero-structure, the free standing monolayer, and of the 7-layer noble metal substrate, respectively. Whereas, the binding energy of $GdAg_2$ on Ag(111) is in the range 3.10-3.97 eV per unit cell in the three stacking configurations, a considerably higher binding energy (between 9.96-10.48 eV) is found for $GdAu_2$ on Au(111). Such energy limits the spontaneous arrangements and orientation of the $GdAu_2$ alloy layer on the Au surface compared with the case of Ag.

The $GdAg_2$ alloy layer can be strained further by depositing Gd at a slightly lower temperature (240ºC). As shown in Figure 4, at this temperature large portions of the surface are tessellated by hexagonal cells of uniform size, formed by a $GdAg_2$ layer with a lattice constant of 5.5Å. Each hexagonal tile is 5-6 $GdAg_2$ unit cells wide (Figure 4b). At this critical size strain cannot be sustained further and discommensuration lines appear as sketched in Figure 4c. Despite the clear change in the STS density of states measured at the disconmensuration lines (Figure 4d), the spectrum measured on the tile closely resembles the one of the moiré superstructures reported in Figure 1. The main difference is a further decreased of the low-energy density of occupied states, where the peak at -440meV remain almost unaffected and the lowest unoccupied electronic feature shift further to +270meV. This agrees with the predicted trend of the onset of the D band as a function of lattice periodicity already discussed for the moiré superstructures.

The magnetic character of the GdAg$_2$ alloys can be probed by considering the energies of the in-plane and out-of-plane magnetic orientations for the three stacking shown in Table 2. The in-plane magnetization is energetically preferred with respect to the out-of-plane direction by 168 meV for hcp and fcc stacking. On the contrary, the top stacking configuration has no anisotropy energy and, is energetically much less stable (0.7-0.8eV). An in-plane ferromagnetic character was experimentally observed in magnetic measurements averaging over the moiré superstructure of GdAg$_2$ [16]. As the continuity of the layer imposes that GdAg$_2$ alloy will be found also in top position, our calculations suggest the formation of a magnetic texture along the layer, where top positions behave as paramagnetic dots embedded in an otherwise in-plane ferromagnetic layer. Even if the magnetic hardness here described were overestimated by the imposition of commensurate boundary conditions, which increases the atomic orbitals overlap of the alloy and of the Ag(111) substrate, we expect that the trend in the magnetic anisotropy will be preserved.

The calculated magnetic anisotropy energies between in-plane and out-of-plane spin orientation are quite large in fcc and hcp, and would correspond to a Curie temperature of 1950K according to mean field theory. The observed critical temperature of 85K [16] can be understood in terms of in-plane magnetic disorder, where the spin orientation is parallel to the surface but fluctuates between different in-plane orientations. The very low out-of-plane anisotropy in the top configuration should further weaken the total magnetic state and contribute to reduce the observed T$_c$. Free standing alloy layers subjected to strain present similar anisotropy energies (see Table S1 of Supporting Information).

## Conclusions

In conclusion, the intimate relation between atomic structure, electronic and magnetic properties allows for controlling through structural relaxation and coupling effects the physical properties of materials and for modulating in space the energy structure and the magnetic order. Through the comparison between the experimental structural and local spectroscopic measurements and the theoretical predictions we survey two major electronic contributions in the occupied and empty band structure of the incommensurate GdAg$_2$ monolayer grown on Ag(111) as a function of in-plane lattice strain. This leads to surface electronic hetero-structuring. Furthermore, out-of-plane structural relaxation and buckling of the atomic structure result in a variation of the interlayer distance, texturing the electronic and magnetic properties of the GdAg$_2$ monolayer with the periodicity of the moiré superstructure. We predict that the magnetic hardness changes across the moiré superstructure leading to a ferromagnetic layer with paramagnetic dots corresponding to the top stacking configuration of the GdAg$_2$ alloy. We believe that the intimate relation found here between structural, electronic and magnetic properties have a general validity in weakly interacting layered systems, most of which form moiré superstructures.

## Acknowledgements

AC and LV acknowledge the financial support of the Spanish ministry of economy (MAT2013-46593-C6-4-P). MJV and BX acknowledge two ARC grants (TheMoTherm #10/15-03 and AIMED # 15/19-09) from the Communauté Française de Belgique, a PDR project (GA T.1077.15) from the Fonds National pour la Recherche Scientifique (Belgium). Computer time was provided by CECI, SEGI, Zenobe/CENAERO (Walloon region GA 1117545), and PRACE-2IP and 3IP (EU FP7 GA RI-283493 and RI-312763) on EPCC Archer.

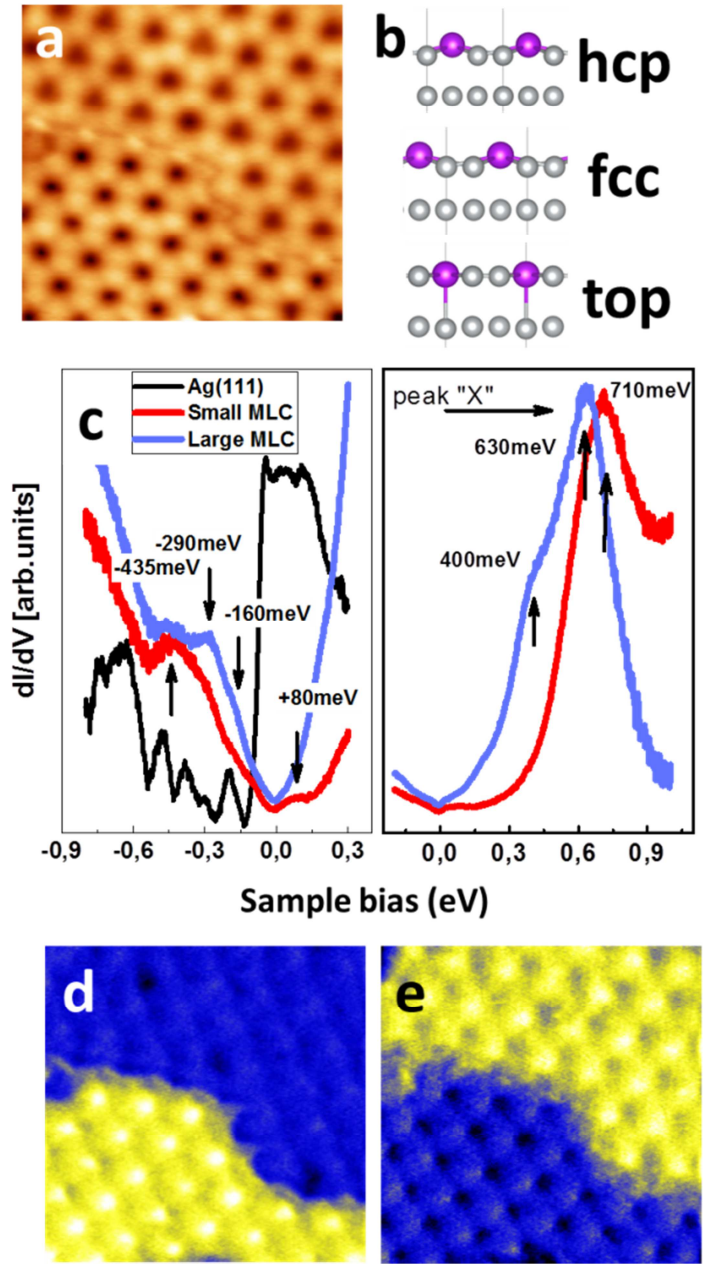

**Figure 1. Structure and electronic properties of a monolayer thick GdAg$_2$ on Ag(111).** (a) Two GdAg$_2$ moiré superstructures differing in lattice constants are seen in topographic images. Up: large moiré lattice constant (LMLC); down: small moiré lattice constant (SMLC). Image size 20nm$^2$ (b) Calculated relaxed geometry for the alloy layer in hcp, fcc and top configuration of one monolayer of GdAg$_2$ on Ag(111). (c) dI/dV spectra acquired on the two moiré superstructures. (d-e) Conductance maps at 380meV and 700meV, showing the localization of the electronic peaks in only one of the LMLC or the SMLC. All experimental data presented here have been acquired using a low temperature scanning tunneling microscope operated at 77K [20]

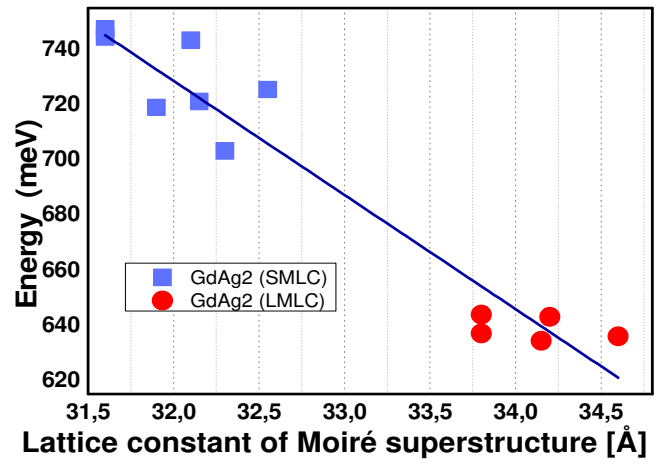

**Figure 2**. **Correlation between the lattice constant of the moiré pattern and the energy of electronic structure.** Energy shifts around equilibrium positions are observed monitoring peak X both in the SMLC (i.e. large alloy unit cell) and LMLC (i.e. small alloy unit cell) patterns (blue and red symbols). The line is a linear fit of the experimental data points.

|  | Experimentally observed | | Values expected [Hermann] | |
|---|---|---|---|---|
| Moiré periodicity | 32Å | 34Å | 32.12±0.1Å | 34.2±0.1Å |
| Rotation angle $\gamma$ | 28±1º | 20±1º | 28.7±0.1º | 20.9±0.1º |
| Atomic distance between Gd atoms | 5.23±0.2Å | 5.13±0.2Å | 5.247±0.001Å | 5.156±0.001Å |
| Nearest neighbor distance |  |  | 3.033Å | 2.981Å |
| $\beta$ |  |  | 3.1±0.2 º | 13.7±0.2º |
| $\alpha$ |  |  | 34.62±0.02º | 34.67±0.02º |

**Table 1**. Comparison between the experimentally observed and the calculated valued of the moiré super-structures using the Hermann model[21].

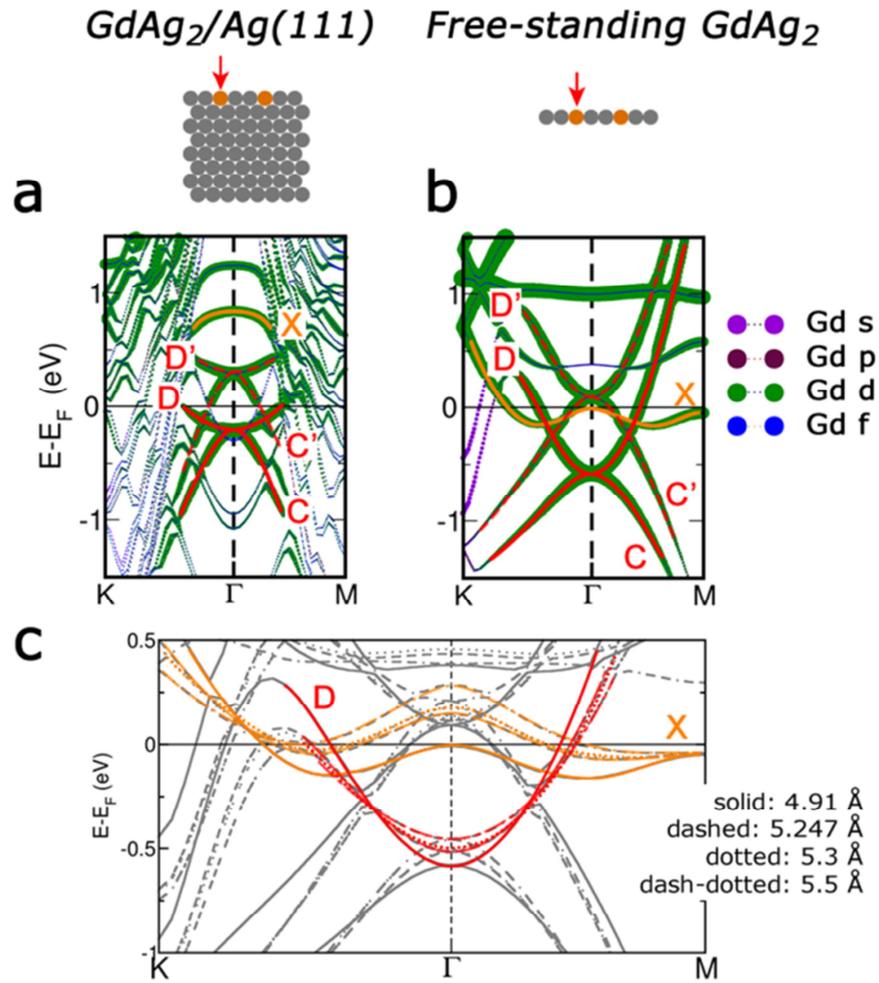

**Figure 3. Density functional theory calculations of the band structure of GdAg$_2$:** (a) Calculated band structures of the GdAg$_2$ alloy layer with (√3x√3)R30º periodicity, in fcc configuration on Ag(111) (left panel). Color lines highlight the band structure responsible for the main observed density of states in Figure 1. (b) Calculated band structure for the free-standing GdAg$_2$ alloy layer with the same lattice constant as the Ag (111) substrate (c) Calculated band structures of the free standing GdAg$_2$ alloy layer for different lattice constants. The main bands that are observed in STS spectroscopy are highlighted.

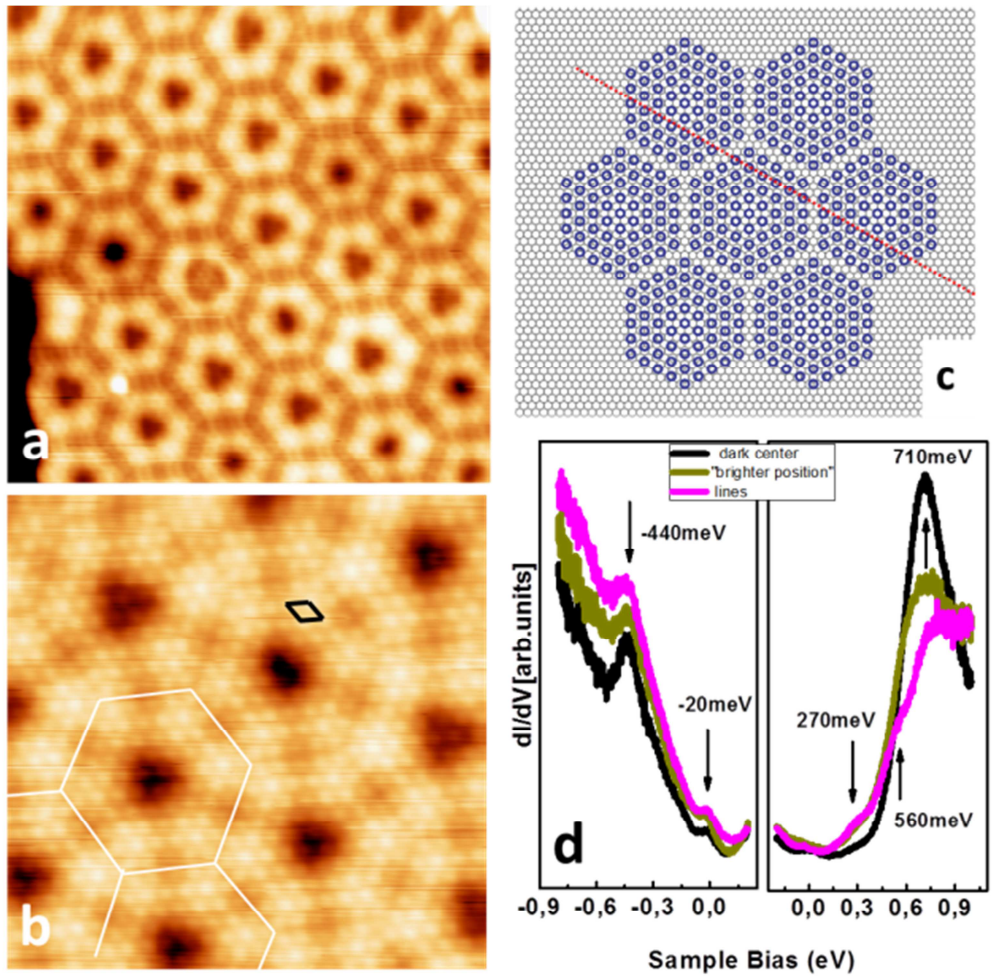

**Figure 4. Hexagonal network of GdAg$_2$ under large strain**. (a-b) Topographic images of the alloy layer showing hexagonal patterns tessellating the Ag(111) surface. (a. 25x25nm, 1eV; b. 13x13nm, -31meV). The alloy unit cell (black) and hexagons (white) are drawn on the figure as guide to the eye. (c) Graphical representation of the hexagonal network. (d) dI/dV spectra taken at different positions moving from the center towards the disconmensutration line.

| Simulated magnetization anisotropy energy | In-plane | | | Out-of-plane | | |
|---|---|---|---|---|---|---|
| Surface alloy position | hcp | fcc | top | hcp | fcc | top |
| Energy relative to in-plane fcc (meV) | 24 | 0 | 852 | 192 | 168 | 852 |

**Table 2**. The calculated out-of-plane magnetic anisotropy ($E_{\text{out-of-plane}} - E_{\text{in-plane}}$) of hcp, fcc and top surface alloy configurations for GdAg$_2$ alloy with Ag substrate. The energies are in meV per formula unit of alloy, and relative to the lowest energy configuration. Note that a (much smaller) in-plane anisotropy exists as well: the out-of-plane anisotropy is a signature of the overall strength of the ferromagnetic state.

# Electronic Supplementary information

# Strain-induced effects in the electronic and spin properties of a monolayer of ferromagnetic GdAg$_2$

*Alexander Correa[1,2,3], Bin Xu[4,5,6], Matthieu J. Verstraete [4,5], Lucia Vitali[1,3,7]*

[1] Departamento de física de materiales, Universidad del País Vasco, 20018 San Sebastian (Spain)

[2] Donostia International Physics Center, 20018 San Sebastian (Spain)

3 Centro de Fisica de Materiales (CSIC-UPV/EHU) y Material Physics Center, 20018 San Sebastian (Spain)

[4] CESAM and Département de Physique, Université de Liège, B-4000 Sart Tilman, Belgium

[5] European Theoretical Spectroscopy Facility (http://www.etsf.eu)

[6] Department of Physics and Institute for Nanoscience and Engineering, University of Arkansas, Fayetteville, Arkansas 72701, USA

[7] Ikerbasque Foundation for Science, 48013 Bilbao (Spain)

## ESI1. GdAg$_2$ lattice structure

The GdAg$_2$ is a stoichiometric alloy, whose structure closely resembles the previously reported GdAu$_2$ [1]. Following this, a model is sketched in Figure ESI1. The topmost layer is a honeycomb lattice (red line) delimited by silver atoms and centered around one gadolinium atom. The orange rhombus delimits the unit cell of the alloy, whose size is given by the distance between the Gd atoms.

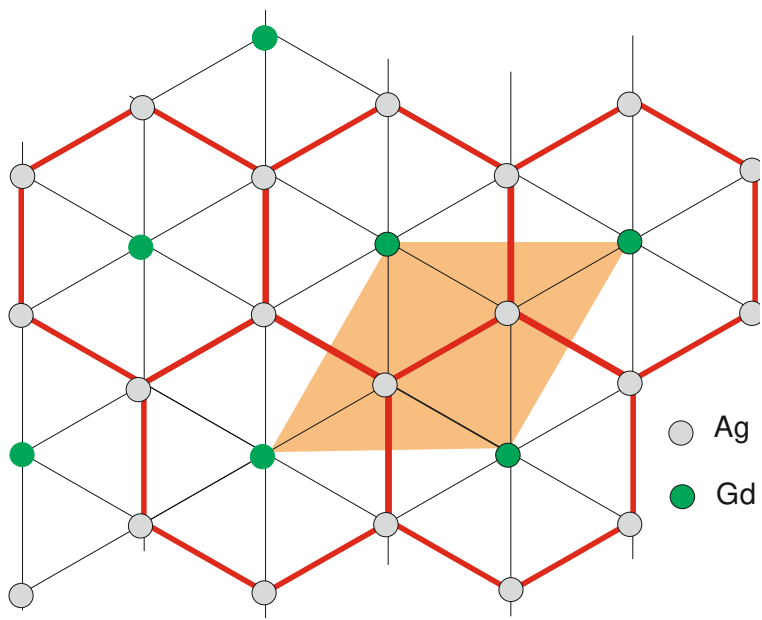

**Figure ESI1. Sketch model of the chemical structure of the GdAg$_2$ alloy**. Green and gray circles represent the gadolinium (Gd) and the silver (Ag) atoms in the layer. The unit cell of the structure, which contains one Gd atom and two Ag atoms is drawn in orange.

In order to form this structure gadolinium atoms are evaporated in ultrahigh vacuum on the Ag(111) previously prepared by cycles of Ar+ ion sputtering and subsequent annealing. The Ag(111) surface held at a temperature between 280º and 320ºC to obtain an ordered GdAg$_2$ alloy. At a lower temperature ~240ºC the hexagonal phase predominantly forms. After the preparation the sample was transferred into the STM where the surface was measured at a temperature of 77K, if not otherwise stated.

**S2. Modeling details**

The electronic structures of GdAg$_2$ alloys are studied by density functional theory (DFT) as implemented in the ABINIT open source package.[2,3] A plane wave basis set (plane wave energy cut off of 20 Hartree) is used with the projector augmented wave (PAW) method[4] to abstract the core states. PAW datasets were generated with the ATOMPAW code[5] (11 valence electrons for Ag, and 18 for Gd). The spin orbit coupling is included in the standard way, perturbatively on the PAW orbitals. The local density approximation (LDA) for the exchange correlation energy was employed. The f-electron states of Gd are extremely localized and strongly correlated: in order to represent them correctly we go beyond (semi)local DFT. Here the LDA+U technique is used,[6] with a Hubbard U parameter of 6.7~eV and J of 0.7~eV, which fixes the Gd f states about 9 eV below the Fermi level. The substrate for the alloy monolayer is modeled as a 7-layer Ag (111) slab with the in-plane lattice constant of the relaxed bulk of Ag (2.83 Å). The alloy layer and two Ag sublayers were relaxed.

The treatment of the incommensurate moiré of the alloy overlayer structure is a crucial point to compare with experiment. The experimental alloy structure has a long-range periodicity, giving rise to a moiré pattern. This structure is simulated in DFT using a commensurate approximant: the √3×√3 supercell cell of the noble metal (111) surface is overlaid with the alloy unit cell. The in-plane lattice constant is fixed to the relaxed bulk DFT value (2.83 Å) for Ag. The strain compared to experiment (imposed by commensurability) is quite large: 8%. The substrate is modeled as a slab of 7 layers of Ag, constructed along the (111) direction. A minimum of 11 Å of vacuum was used to separate the slabs, in order to avoid interference between periodic images. The structures were relaxed for the alloy and the top three layers of substrate atoms, until forces were below $2\times10^{-3}$ Ha/bohr. In several instances, we checked that the differences in the electronic properties (band structure) are negligible compared to structures converged to maximum forces of less than $10^{-5}$ Ha/bohr.

The incommensurability of the experimental primitive surface unit cells implies that, along the moiré pattern, the alloy is centered at different positions of the √3×√3 substrate. We study the three representative shifts through which the system must pass, where the alloy atoms fall either in the "natural" fcc positions, in the hcp positions with respect to the last two layers of substrate, or on top of surface atoms (top). The three will recur periodically in the moiré pattern. We will see below that fcc and hcp share most properties but that the top position is quite different in bonding and electronic structure. Because of the lattice mismatch and strain, the electronic bands of the surface alloy and subsurface noble metal layer are shifted in energy. Based on tests with variable lattice parameters, the shifts are not homogeneous for all of the bands linked to a given layer. Further, the charge transfer between the alloy and surface will also depend on the strain, and will give a relative shift of the alloy bands with respect to the bulk noble metal bands. Finally, it is well known that the unoccupied Kohn-Sham energies are not quantitatively accurate predictions of band positions. As a result, the comparison of bands with experimental STS features in the following will be qualitative, and used to explain their nature, relative positions, and dependency on alloy position and magnetization.

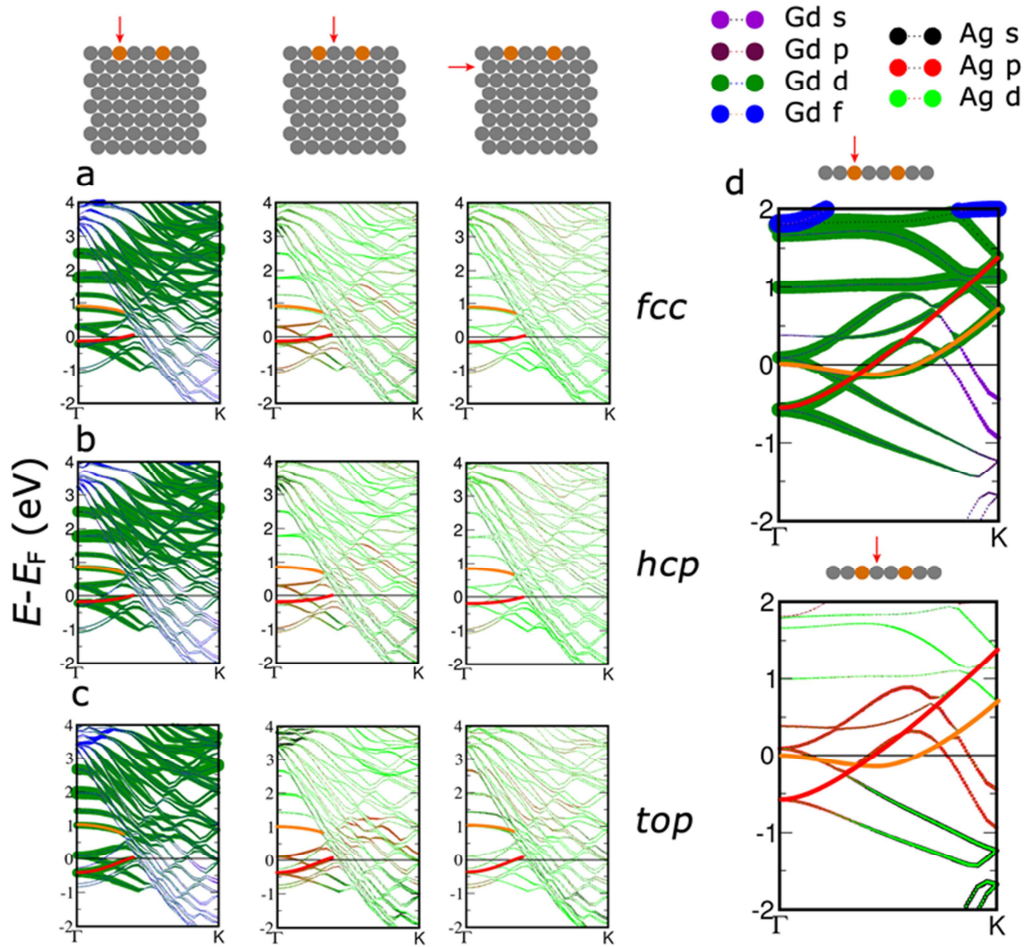

**Figure ESI2. Comparison of the electronic band structures between GdAg$_2$ alloy on Ag substrate and free standing GdAg$_2$ monolayer with the theoretically relaxed lattice constant.** (a-c). Calculated band structures of alloy with substrate for fcc, hcp, and top configurations. Colors and line width denote the contributions from different orbitals of Gd and Ag atoms (in the alloy layer and the first sublayer). (d) Calculated band structures of the free-standing alloy monolayer. The magnetization is in-plane.

Figure ESI2 compares the electronic states for different layer stackings (hcp, fcc, top) with those of the free-standing alloy layer. In each case, the atomic orbital projections

on Gd and Ag (as well as the first layer of substrate) are shown, and the two bands visible in STM/STS are highlighted in orange and red.

Figure ESI3 shows the weak effect of magnetic moment orientation on the electron band structure, in the three stackings. All cases are topologically similar, with a shift down in bands near the Fermi level for the top stacking. Some band splittings can be observed in the out-of-plane case, e.g. around 2eV at the Gamma point for top stacking.

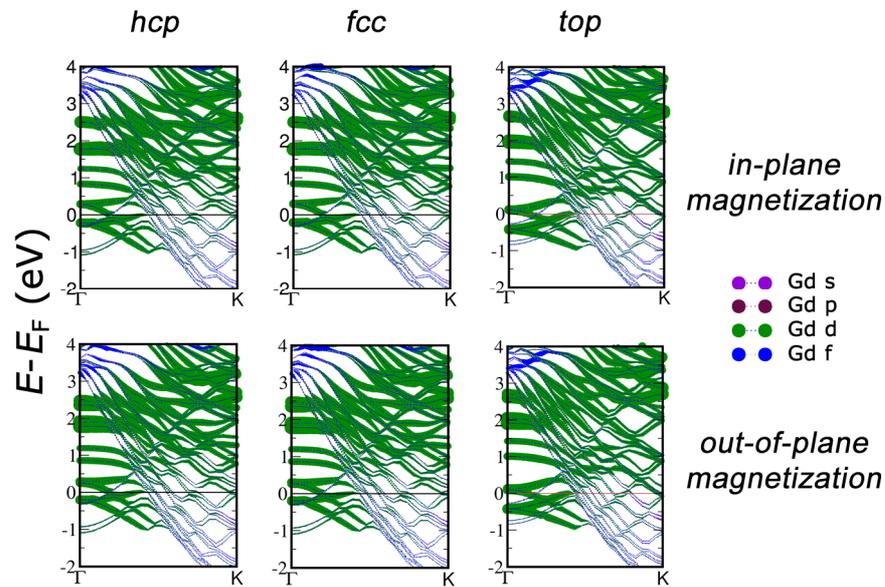

**Figure ESI3. Calculated band structures for three relative positions of the GdAg$_2$ alloy with respect to the substrate, including spin-orbit coupling.** Three columns from left to right are hcp, fcc, and top configurations. The top row is for in-plane magnetization, and the bottom row is for out-of-plane magnetization.

The Gd-d orbital character of the electronic bands near the Fermi level is shown in Figure ESI4 for the free standing alloy layer. Each pair of spin-orbit split bands has a specific orbital character, with the crucial bands near $E_F$ being in-plane $d_{xy}$ and $d_{x^2-y^2}$, as could be expected from hybridization arguments with the Ag in plane. The differentiation between $d_{yz}$ and $d_{xz}$ orbitals is due to the hexagonal lattice symmetry, which breaks their equivalence.

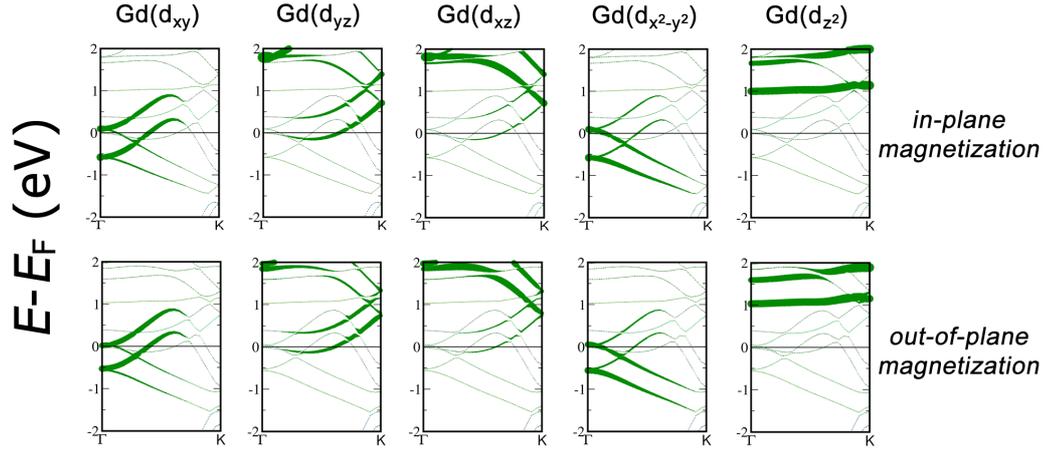

**Figure ESI4. Calculated Gd 5*d* orbital contributions in the GdAg$_2$ free standing alloy layer, for in-plane and out-of-plane magnetization.**

In Table ESI1 we show the magnetic anisotropy of isolated monolayers with changing strain conditions. The order remains the same as in the supported case, and the anisotropy is almost independent of strain.

| Magnetization | $E_{\text{out-of-plane}}$ - $E_{\text{in-plane}}$ | | | |
|---|---|---|---|---|
| Lattice constant (Å) | 4.91 | 5.247 | 5.3 | 5.5 |
| Anisotropy (meV) | 157 | 175 | 180 | 186 |

**Table ESI1**. **The calculated out-of-plane magnetic anisotropy ($E_{\text{out-of-plane}} - E_{\text{in-plane}}$) of the free standing monolayer alloy, with respect to lattice constant.** The energies are in meV per formula unit of alloy, and relative to the lowest energy configuration for each case. Note that much smaller in-plane anisotropy exists as well, and determines the Curie temperature of the system: the out-of-plane anisotropy is a signature of the overall magnetic strength of the FM state. As with the supported case in Table 2 of the main text, the easy axis is in-plane.

**ESI3. Density of states on GdAg2 superstructure**

In Figure ESI5, the density of states that characterizes the apparent hills and valleys of the GdAg$_2$ moiré superstructure (in panel S5a as red and green line, respectively) is shown. The localization of the spectroscopic features is evident by comparing topographic image and energy maps measured at the energy of the peaks observed in the dI/dV spectra.

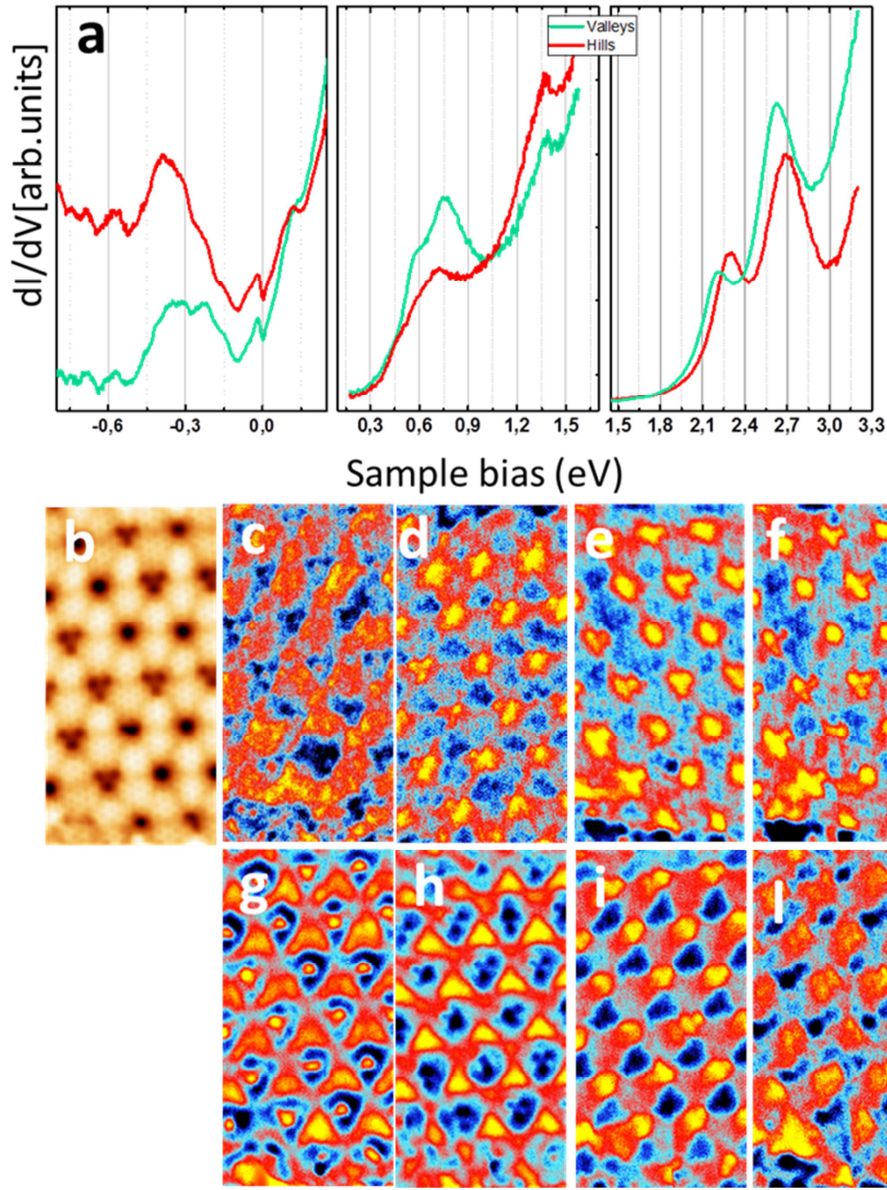

**Figure ESI5**. Topographic image and the density of states measured on different position of the GdAg2 moiré superstructure at the temperature of 1K (a) dI/dV spectra on hills and valley position of the moiré superstructure (b) measured at (c). -300meV, (d) -200meV; (e) 600meV, (f) 790meV, (g) 2.2eV; (h) 2.3eV; (i) 2.6eV; (l) 2.7eV.

## ESI4. Density of states on the "hills" position of the moiré patterns

A comparison of the density of states measured on the two moiré patterns shown in Figure 1 of the main text (hills positions) is reported in Figure ESI6. The spectra are shown in separate panels for a better visualization. Spectroscopic differences between two moiré patterns are found in the whole range of density of states. These spectral features here observed characterize each of the two moiré pattern as evinced by the energy maps achieved on the topographic image (panels b-h)

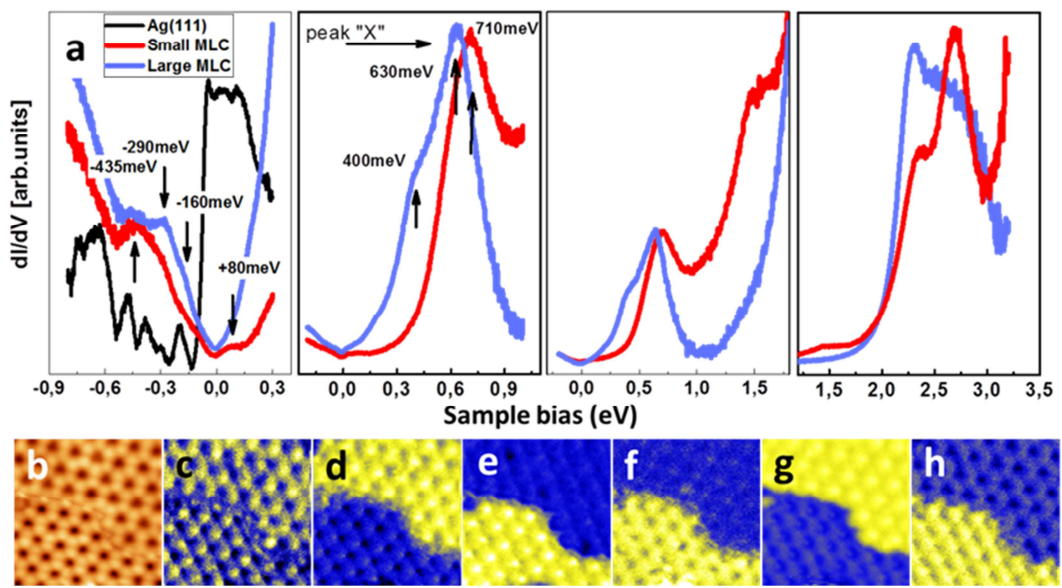

**Figure ESI6. The LDOS and conductance maps measured on the "hill" position of two differing moiré super-structures of $GdAg_2$.** (a) dI/dV spectra. (b) Topographic image. (c-h) Conductance maps at the energy corresponding to the features in the density of states (c). -280meV, (d). 380meV, (e). 700meV; (f). 1.4eV; (g). 2.2 eV; (h). 2.7eV.

## ESI5. Density of states on the "dark" position of the moiré patterns

The local densities of states measured on the "valley" positions of the two moiré patters are reported in Figure ESI6.

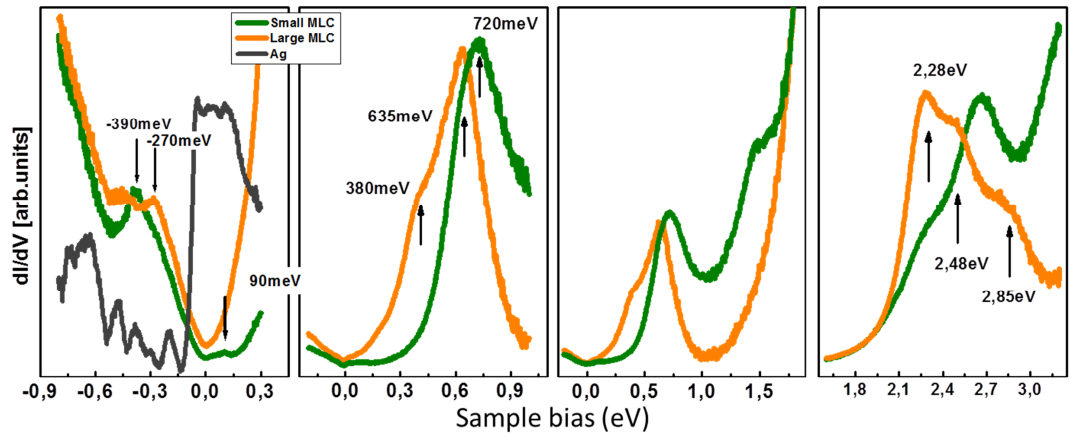

**Figure ESI7. LDOS measured on the valley positions of the moiré superstructures of GdAg$_2$. (a)** dI/dV spectra have been achieved on the small and large moiré super-lattice constants (SMLC, LMLC)

## ESI6. Moiré superstructures of GdAg$_2$ on Ag(111)

High resolution image of two moiré superstructures of the GdAg$_2$ monolayer alloy grown on Ag(111) are shown in Figure ESI8. The white lines highlight that these differ in periodicity and relative orientation. Atomically resolved images of the two superstructures (panel b and c) show that the alloy unit cell (small black rhombus) is also rotated of an angle γ with respect to the unit cell of the superstructures (white rhombus). The lattice constant of the GdAg$_2$ unit cell, 5.15±0.05Å and 5.25±0.05Å, and the angle of rotation γ characterize these structures.

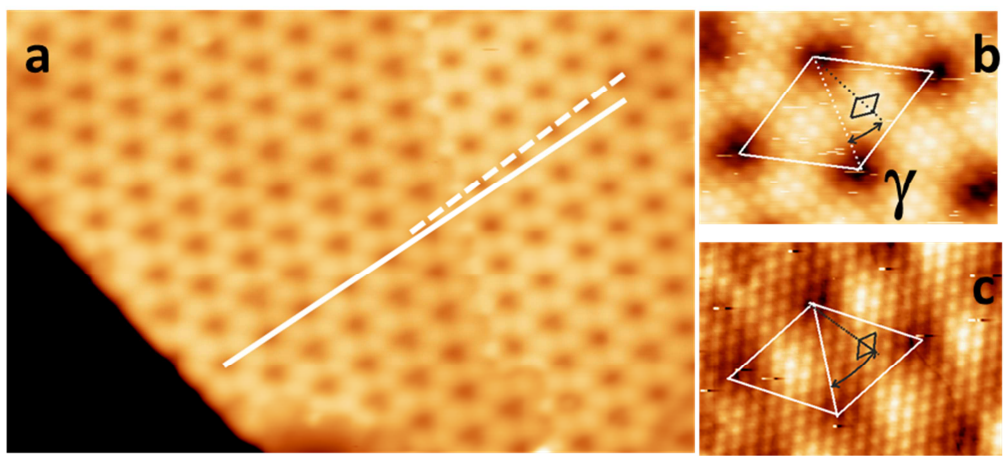

**Figure ESI8. GdAg$_2$ moiré structures on Ag(111).** **(a)** Topographic image of one monolayer of GdAg$_2$ alloy on Ag(111) showing two moiré structures (230x150Å). The white lines underline the relative rotation of the two moiré patterns. (b and c) Atomic resolution of the two moiré patterns achieved on the left and on the right superstructure of panel a, respectively. The white and black rhombi show that GdAg$_2$ lattice has a different orientation on the two moiré patterns.

## ESI7 Lattice constant estimation

Moiré superstructures are envisioned as the superposition of two incommensurate layers, as for example an overlayer and a supporting substrate, and/or by their relative rotations. Each combination of different lattice constants and angle of rotation leads to a moiré superstructure with different characteristics as periodicity and angle of rotation. The lattice constant, the relative orientation of the overlayer, in the present case of GdAg$_2$ alloy, can be calculated using the coincidence model described by K.Hermann[7] knowing the parameters of the substrates Ag(111) and/ or of the moiré superstructures.

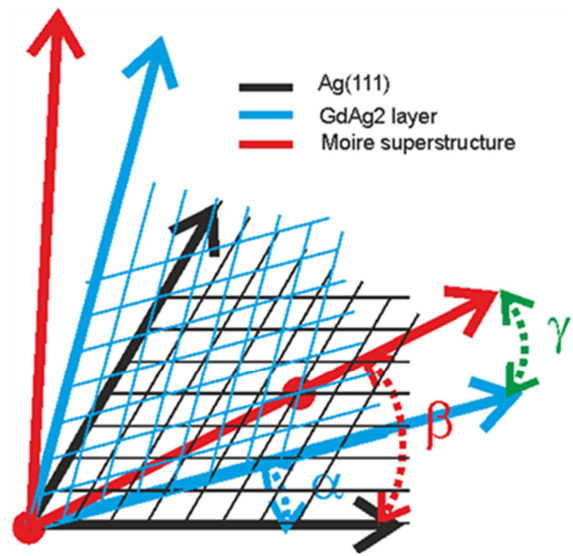

**Figure ESI9 Scheme of the coincidence lattice model.** The alloy lattice (blue) is superimposed on the substrate (black) and generates a coincidence networks (red) that forms the moiré pattern.

The lattice of GdAg$_2$ (blue) and Ag(111) (black) differs in unit cell size and form a relative angle of rotation α. A coincidence network forms by the atoms of the alloy layer superposing on atoms of the Ag(111) substrate. This defines the orientation and lattice constant of the moiré superstructure which is characterized by an angle β with respect to the substrate and an angle γ with respect to the alloy layer. The experimentally observed angle γ is shown in Figure ESI8.

According to this model, the periodicity of the superstructure follows the following relation:

1.

$$M = \frac{a}{\sqrt{\left(1 + \left(\frac{a}{b}\right) * \left(\frac{a}{b}\right) - 2\left(\frac{a}{b}\right)\cos(\alpha)\right)}}$$

where *a* and *b* are respectively the lattice constants of Ag(111) and the first nearest neighbor distance in the alloy layer. The angle α defines the relative orientation between the alloy layer and Ag(111). The angle β between the moiré superstructure and the supporting substrate can be calculated as follows:

2. $\beta = \arccos\left(\dfrac{\cos(\alpha) - \left(\frac{a}{b}\right)}{\sqrt{1 + \left(\frac{a}{b}\right)*\left(\frac{a}{b}\right) - 2\left(\frac{a}{b}\right)\cos(\alpha)}}\right)$

The angle γ is the angle between the moiré superstructure and the alloy layer and can be related to α and β by the simple equation:

3. γ=β−α

Using this model, fixing only the substrate lattice constant to 2.9Å, we have calculated the possible combinations of first atom nearest neighbor distance *b* in the alloy and relative rotation angle α with respect to Ag(111) which lead to the periodicities of moiré superstructures experimentally observed. Among the possible parameters calculated for this system we report in Table 1 the ones that reproduce the experimentally observed superstructure.

|  | Experimentally observed | | Values expected[1] | |
|---|---|---|---|---|
| moiré periodicity *M* | 32Å | 34Å | 32.12±0.1Å | 34.2±0.1Å |
| Rotation angle γ | 28±1° | 20±1° | 28.7±0.1° | 20.9±0.1° |
| Atomic distance between Gd atoms | 5.23±0.2Å | 5.13±0.2Å | 5.247±0.001Å | 5.156±0.001Å |
| First nearest neighbor's distance |  |  | 3.033Å | 2.981Å |
| Rotation angle β |  |  | 3.1±0.2 ° | 13.7±0.2° |
| Rotation angle α |  |  | 34.6° | 34.65° |

**Table ESI2.** Comparison between the experimentally observed values of the moiré super-structures and those calculated using the Hermann model.